\documentclass[aps,prl,reprint,superscriptaddress]{revtex4-2}

\usepackage{stackengine,graphicx,xcolor}
\usepackage{amsmath}
\usepackage{etaremune}
\usepackage{braket}
\usepackage{mathtools}
\usepackage{anyfontsize}
\usepackage{bm}
\usepackage{hyperref}

\usepackage[T1]{fontenc}
\usepackage{amssymb}

\def\CircleArrowleft{\ensuremath{%
  \reflectbox{\rotatebox[origin=c]{180}{$\circlearrowleft$}}}}
\def\CircleArrowright{\ensuremath{%
  \reflectbox{\rotatebox[origin=c]{180}{$\circlearrowright$}}}}

\begin{document}
\title{Ultrastrong magnetic light-matter interaction with cavity mode engineering}
\author{Hyeongrak Choi}
\email[]{choihr@mit.edu}
\author{Dirk Englund}
\email[]{englund@mit.edu}
\affiliation{Research Laboratory of Electronics, Massachusetts Institute of Technology, Cambridge, Massachusetts 02139, USA}

\begin{abstract}
\textbf{Abstract}: Magnetic interaction between photons and dipoles is essential in electronics, sensing, spectroscopy, and quantum computing. However, its weak strength often requires resonators to confine and store the photons. Here, we present mode engineering techniques to create resonators with ultrasmall mode volume and ultrahigh quality factor. In particular, we show that it is possible to achieve an arbitrarily small mode volume only limited by materials or fabrication with minimal quality-factor degradation. We compare mode-engineered cavities in a trade-off space and show that the magnetic interaction can be strengthened more than $10^{16}$ times compared to free space. Proof-of-principles experiments using an ensemble of diamond nitrogen-vacancy spins show good agreement with our theoretical predictions. These methods enable new applications from high-cooperativity microwave-spin coupling in quantum computing or compact electron paramagnetic resonance sensors to fundamental science such as dark matter searches.
\end{abstract}

\maketitle

\section{Introduction}

\noindent The magnetic light-matter interaction has a wide range of applications including quantum computing \cite{vandersypen2005nmr} and sensing \cite{crescini2020cavity, mohammady2018low}, nuclear magnetic resonance (NMR) \cite{bloembergen1948relaxation}, active microwave (MW) devices \cite{pozar2011microwave}, and MW amplification by stimulated emission of radiation (maser) \cite{oxborrow2012room}. However, the magnetic interaction is intrinsically weak compared to the electrical counterpart \cite{novotny2012principles}. It is possible to increase the coupling by designing resonators with a strong magnetic field and long interaction time, i.e., a small magnetic mode volume ($V_B$) and a high quality factor ($Q$). This is especially important in quantum applications requiring a strong interaction between a single photon and magnetic dipoles. 

\begin{figure*}
    \centering
    \includegraphics[width=\textwidth,trim=15 15 4 16,clip]{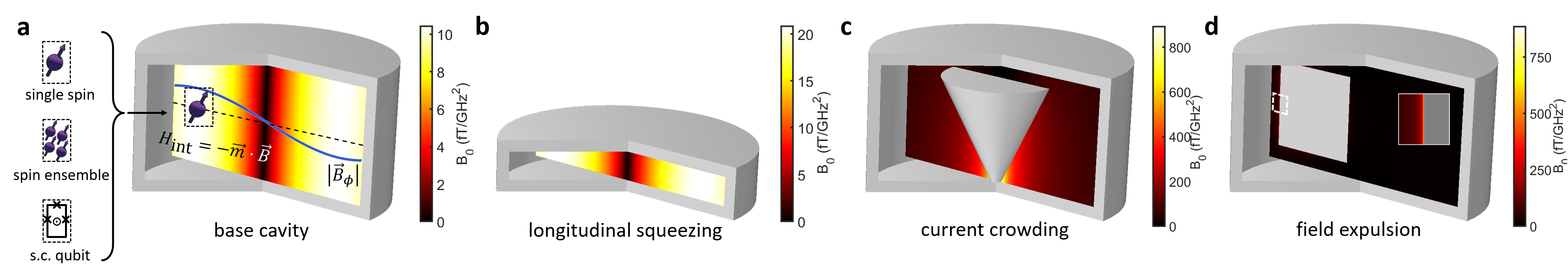}
    \caption{\textbf{Magnetic cavity-QED and mode volume engineering.} \textbf{a} We consider hollow cylindrical resonators that can support non-TEM modes. Magnetic materials, such as single spin, spin ensemble (magnon) or superconducting qubits with magnetic transition dipole moment ($\vec{m}$) can be placed at the magnetic anti-node, for maximum interaction. The magnetic interaction Hamiltonian is $H_\text{int} = -\vec{\textbf{\text{m}}}\cdot\vec{\textbf{\text{B}}}$. \textbf{b} Decreasing the height of the cavity (longitudinal squeezing) enhances the magnetic field inversely proportional to the square root of the height. \textbf{c} Current crowding in the re-entrance strengthens the magnetic field. \textbf{d} Thin metallic structure with near unity demagnetization factor produces magnetic hot spots by field expulsion, analogous to Meissner's effect.}
\end{figure*}

A magnetic dipole interacts with the EM fields through the interaction Hamiltonian $H_{\text{int}} = -{\vec{\textbf{\text{m}}}}\cdot{\vec{\textbf{\text{B}}}}$ (Fig.~1\textbf{a}), where ${\vec{\textbf{\text{m}}}}$ is the magnetic transition dipole moment and ${\vec{\textbf{\text{B}}}}$ is the magnetic flux density. (For simplicity, we will refer to $\vec{\textbf{\text{B}}}$ as the magnetic field.) For cavity photons, the single-photon magnetic field $B_s$ is
\begin{equation}
    B_s(\vec{\textbf{\text{r}}}) = \sqrt{\frac{hf}{2V_B(\vec{\textbf{\text{r}}})/\mu}},
\end{equation}
where $hf$ is the photon energy, $\mu$ is the permeability of the cavity filling, and $V_B$ is the effective magnetic mode volume of a cavity:
\begin{equation}\label{modeVolume}
    V_B(\vec{\textbf{\text{r}}_\text{e}}) = \frac{\int dV |\vec{\textbf{\text{B}}}(\vec{\textbf{\text{r}}})|^2/\mu(\vec{\textbf{\text{r}}})}{|\vec{\textbf{\text{B}}}(\vec{\textbf{\text{r}}_{\text{e}}})|^2/\mu(\vec{\textbf{\text{r}}_{\text{e}}})},
\end{equation}
where $\vec{\textbf{\text{r}}}_\text{e}$ is the position vector of a dipole and is chosen for maximum $|\vec{\textbf{\text{B}}}(\vec{\textbf{\text{r}}_{\text{e}}})|$. On the other hand, cavity photons have a finite lifetime proportional to the quality factor $Q$, i.e. $\tau = Q/2\pi f$. 
Thus, small $V_B$ and high $Q$ are desirable for light-matter applications.

There have been extensive studies on superconducting qubit \cite{wallraff2004strong,krantz2019quantum} and solid-state spin qubit technologies \cite{kubo2010strong,kubo2011hybrid,kubo2012electron} coupled with coplanar waveguide (CPW) resonators. In CPW resonators, two or more electrically isolated conductors support transverse electromagnetic (TEM) modes with zero cutoff frequency \cite{pozar2011microwave}. The transversal Laplace equation allows two conductors to be proximate for a small mode volume. However, the dielectric loss of substrates limits the $Q$ factor to less than one million, presenting a critical problem in applications that require a long photon lifetime \cite{sage2011study}. On the other hand, recent progress in 3D transmon \cite{paik2011observation} based on hollow metallic cavities achieved high $Q$ over a million \cite{campagne2020quantum}. However, their magnetic mode volume is too large to strongly couple photons with qubits because they can support only transverse electric (TE) and transverse magnetic (TM) modes.

\begin{figure}[b]
    \centering
    \includegraphics[width=\columnwidth,trim=20 15 4 16,clip]{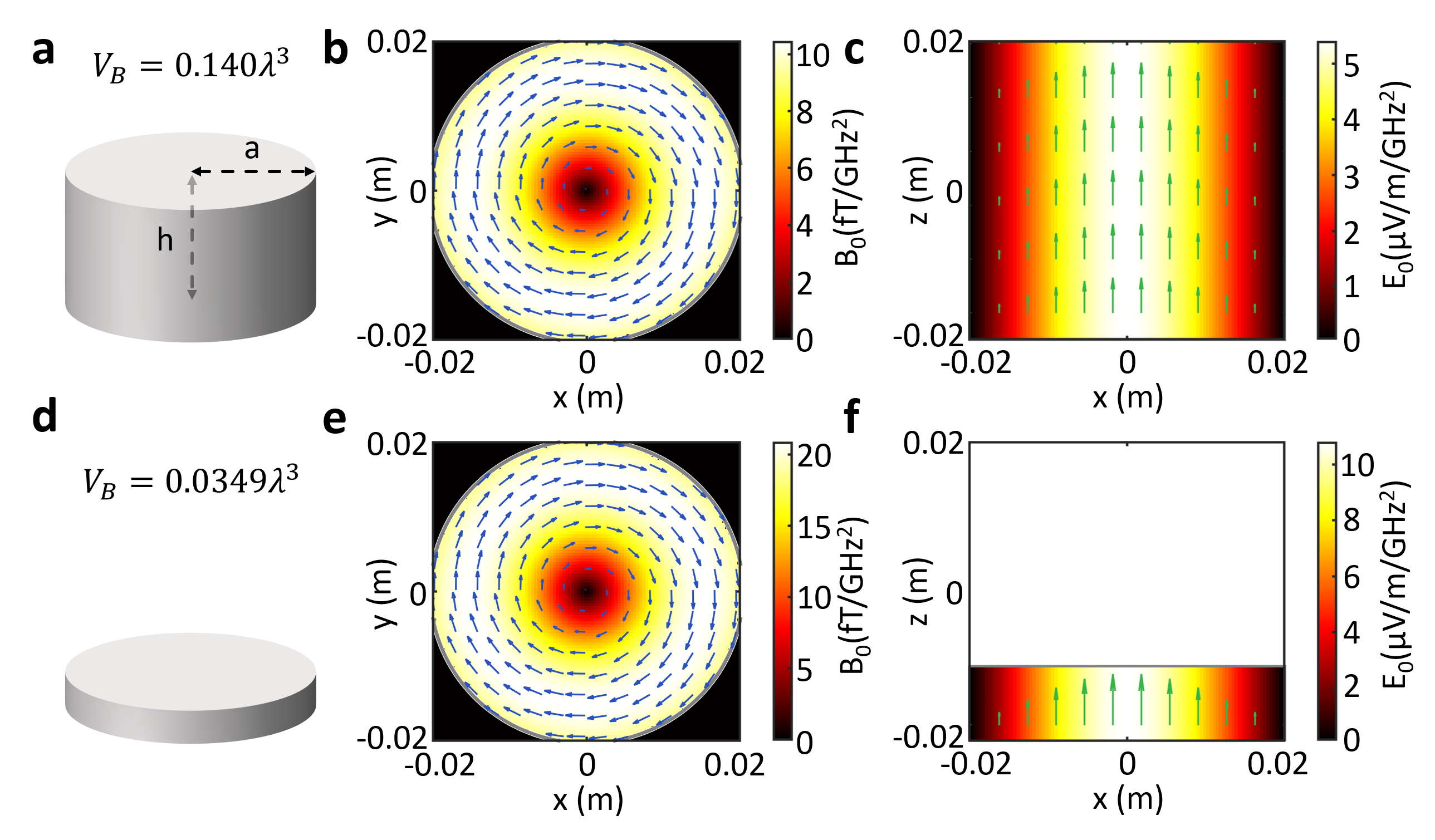}\label{longitudinalSqueezing}
    \caption{\textbf{Longitudinal squeezing of the TM$_{010}$ mode in a cylindrical cavity.} The radii of the cavities are $a=2$~cm, and the heights are \textbf{a}, $h=2$~cm and \textbf{d}, $5$~mm. \textbf{b, e}: magnetic field of TM$_{010}$ mode in $xy$-plane. The field is independent of $z$. The lengths of the arrows are proportional to the field strength. \textbf{c, f}: electric field of TM$_{010}$ mode in $xz$-plane. The mode is azimuthally symmetric. Resonant frequencies of cavities are the same, $f = 5.74$~GHz. The magnetic and electric fields are out of phase.}
\end{figure}

\begin{figure*}
    \centering
    \includegraphics[width=\textwidth,trim=12 4 4 10,clip]{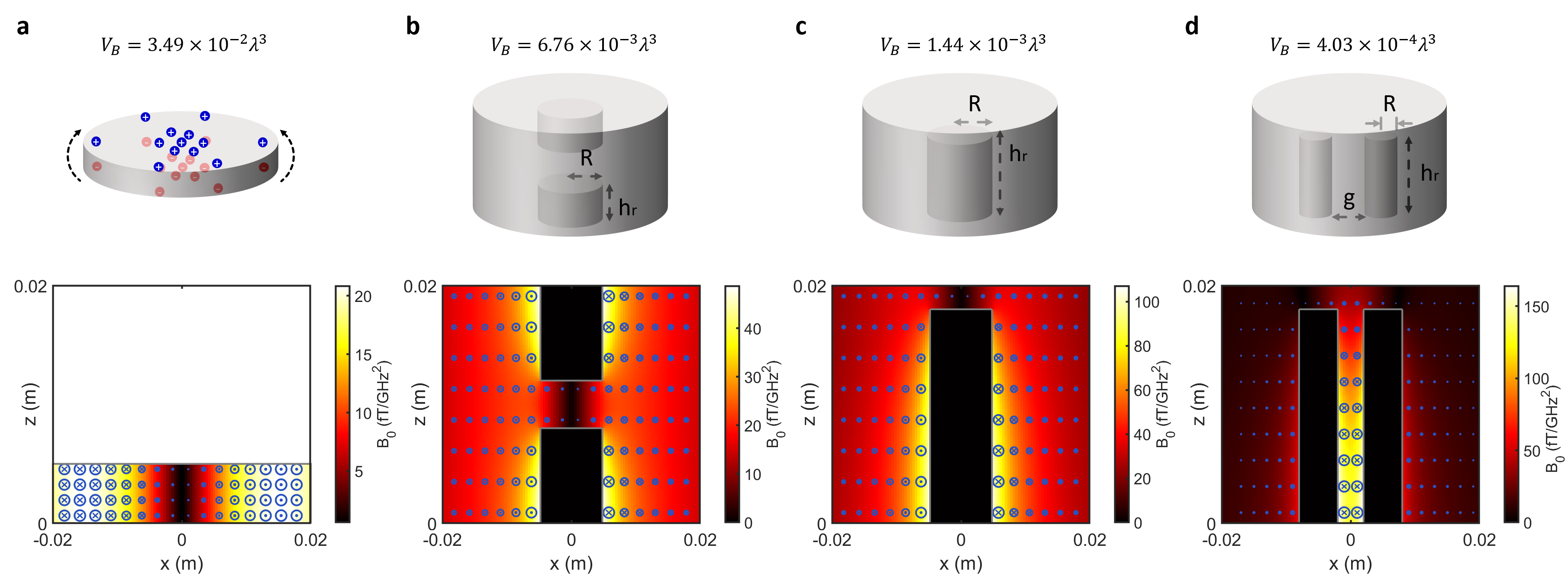}
    \caption{\textbf{Local field enhancement with current engineering.} The colormap represents the norm of the frequency-normalized single-photon magnetic field ($B_0$), and blue circles indicate an out-of-plane magnetic field. The radius of circles is proportional to the magnetic field component in each plot, and $\bullet$ and x inside the circle represents directions. \textbf{a}, Longitudinally squeezed TM$_{010}$ mode. The blue and red circles illustrate charge densities on top and bottom, and the black arrows show out-of-phase current. \textbf{b}, Double split mode cavity ($R=5$~mm). \textbf{c}, Reentrant cavity ($R=5$~mm, $h_r=1.8$~cm). \textbf{d}, Double reentrant cavity ($R=3$~mm, $h_r=1.8$~cm, $g=4$~mm). The currents in the two reentrances are opposite directions. The resonance frequency of each cavity is, \textbf{a}, $f = 5.74$~GHz, \textbf{b}, $f = 3.39$~GHz, \textbf{c}, $f = 2.23$~GHz, and \textbf{d}, $f = 3.21$~GHz.}
\end{figure*}

Here, we present and compare three approaches to mode engineering of hollow metallic cavities for a strong magnetic light-matter interaction: longitudinal mode squeezing, current engineering, and magnetic field expulsion. (i) We first study the longitudinally squeezed TM$_{mn0}$ mode (Fig.~1\textbf{b}). (ii) Because the mode volume is inversely proportional to the magnetic field at one point ($\vec{\textbf{\text{B}}}(\vec{\textbf{\text{r}}_\text{e}})$), we locally increase the magnetic field using current engineering. The framework integrates and derives previously proposed cavity designs such as split mode, loop-gap, and reentrant cavities \cite{le2016towards,carter2007calculation,park2016randomized,creedon2015strong,goryachev2014high}. Furthermore, our mode engineering allows for further reduction of mode volume through current crowding (Fig.~1\textbf{c}). (iii) Finally, motivated by the Meissner effect, we propose a field enhancement method by field expulsion (Fig.~1\textbf{d}). We show that all three methods enable arbitrarily small $V_B$ only limited by the electromagnetic (EM) penetration depth, typically on the order of tens of nanometers for superconductors. The reduction of $V_B$ accompanies a reduction of $Q$, but $Q/V_B$, a figure of merit for light-matter interaction, can increase more than five orders of magnitude than existing designs of cavities for strong interaction; thus, we coined the term ultrastrong interaction. Our proof-of-principles experiments with diamond nitrogen vacancy centers show quantitative agreement with the theoretical predictions. Ultrastrong magnetic interaction by mode engineering enables new applications including microwave-to-optical transduction, low-noise masers, magnetic strong coupling of spin qubits or superconducting qubits, and cavity magnonics.

\section{Results}
\vspace{6mm}
\noindent\textbf{TM$_{mn0}$ mode and longitudinal squeezing} 

\noindent Consider the EM field of the TM$_{mnp}$ modes in the cylindrical cavity (Fig. 2) of radius $a$ and height $h$;
\begin{equation}
\begin{split}
    E_z &= E_0\cdot J_m(k_c r)\cdot \cos(m\phi)\cdot \cos(\frac{\pi p z}{h})\cdot e^{i\omega_{mnp}t} \\
    B_\phi &= E_0\cdot \frac{i}{\omega} k_c\cdot J'_m(k_c r)\cdot \cos(m\phi)\cdot \cos(\frac{\pi p z}{h})\cdot e^{i\omega_{mnp}t},  
\end{split}
\end{equation}
where $k_c$ is the cutoff wave vector of the circular base waveguide, $J_m(r)$ the Bessel function of the first kind with the order $m$, $\omega_{mnp}=c\sqrt{(j_{mn}/a)^2+(p\pi/h)^2}$, and $j_{mn}$ the $n^{\text{th}}$ zero of $J_m(r)$. $m$, $n$, and $p$ represent azimuthal, radial, and longitudinal wavenumbers.

Cylindrical cavities support TM$_{mn0}$ modes where the field is longitudinally uniform. We focus specifically on the lowest-order mode, TM$_{010}$;
\begin{equation}
\begin{split}
    \vec{\textbf{\text{E}}}(\vec{\textbf{\text{r}}}) &= \hat{\textbf{\text{z}}}E_0\cdot J_0(j_{01}r/a)\cdot e^{i\omega_{010}t} \\
    \vec{\textbf{\text{B}}}(\vec{\textbf{\text{r}}}) &= \hat{\bm{\phi}} E_0\cdot \frac{i}{\omega}\frac{j_{01}}{a}\cdot J'_0(j_{01}r/a)\cdot e^{i\omega_{010}t},
\end{split}
\end{equation}
and $\omega_{010}=c\cdot(j_{01}/a)$. The resonant frequency is independent of $h$, and the cavity can be squeezed longitudinally. 

Figures 2\textbf{a}-\textbf{c} show the magnetic and electric field of the TM$_{010}$ mode of the cylindrical cavity with a radius and height of $0.02$~m. We used the finite element method (FEM) with commercial software (COMSOL). (See Methods for the simulation details). We plot the single-photon magnetic field normalized by frequency square, $B_0 = B_s/f^2$. The scale invariance of Maxwell's equations implies that any cavity modes may be scaled with the wavelength. Thus, $V_B$ of the same mode is proportional to the wavelength cube. Equation~(1) shows that $B_s$ is proportional to $f^2$, and $B_0$ serves as a frequency-independent metric determined by structural design.

Figures 2\textbf{d}-\textbf{f} show the mode of the squeezed cavity ($h = 0.005$~m). The cavities in Figs. 2\textbf{a} and 2\textbf{d} have the same resonance frequency, $f = 5.74$~GHz. Squeezing from $h = 0.02$~m to $h = 0.005$~m does not change the mode field profile in the $xy$-plane (independent of $z$), but decreases the volume four times, from $0.140\lambda^3$ to $0.0349\lambda^3$ where $\lambda$ is the vacuum wavelength at resonance. The four-fold increase in the energy density for fixed energy results in a twice stronger magnetic field. We emphasize that TM$_{010}$-like mode generally exists in any metallic hollow cavity. 

TM$_{010}$ mode volume of a cylindrical cavity is (see Methods),
\begin{equation} \label{modeVolume010}
\begin{split}
    V_B &= \frac{0.135}{2\pi}\cdot \frac{j_{01}^2}{|J_{0,\text{max}}'|^2}\cdot h \cdot \lambda^2 \\
    &\approx 0.366 \cdot \frac{h}{\lambda} \cdot \lambda^3,
\end{split}
\end{equation}
where $|J_{0,\text{max}}'| \approx 0.582$ is the maximum absolute value of the derivative of the Bessel function. Equation (\ref{modeVolume010}) shows that $V_B$ can be much smaller than $\lambda^3$ for $h\ll\lambda$. For infinitesimally short cavities, $h\rightarrow 2\delta$ and $V_B \rightarrow 0.73 \delta \lambda^2$ where $\delta$ is the penetration depth of the metal. 

This contrasts with the usual intuition that the mode volume is diffraction-limited ($V_B\gtrsim\lambda^3$). The EM-boundary condition allows the perpendicular electric (parallel magnetic) field to be discontinuous with the surface charge (current) on conductors. Because we are not confining EM fields by diffraction, the diffraction limit does not apply. Note that the EM field only occupies a homogeneous dielectric medium with uniform permittivity and permeability. Thus, our methods also differ from the dielectric approaches \cite{choi2017self,hu2016design,choi2019cascaded}.

\vspace{6mm}
\noindent\textbf{Advantages of superconducting materials} 

\noindent Longitudinally squeezed cavities have $V_B$ only limited by penetration depth. Superconductors have a small penetration depth even at very low frequencies and enable an ultrasmall mode volume. For comparison, superconducting niobium (Nb) has a penetration depth of $40$~nm, while the skin depth of copper (Cu) is $2~\mu$m ($65~\mu$m) at $1$ GHz ($1$ MHz).

As $h$ decreases, a larger fraction of the EM energy is in the conductor, resulting in ohmic losses. Superconductors mitigate losses with much lower surface resistance ($R_s$). Briefly, $R_s\rightarrow0$ as $T\rightarrow 0$ K for BCS superconductors. In practice, $R_s$ is limited by residual resistances caused by trapped flux, impurities, or grain boundaries \cite{tinkham2004introduction}. Overall, Nb has $R_s = 1\sim 10$~n$\Omega$ \cite{palmer1988surface}, while Cu has $R_s \sim 1$~m$\Omega$ \cite{calatroni2019cryogenic} ($\sim 1$ K). Thus, superconductors enable $10^5\sim 10^6$ times smaller loss (larger $Q$ factor).

When superconducting material is used, it is important to ensure that the resonance frequency of the cavities is lower than the gap frequency (Nb: 725 GHz). If the frequency is above or near the gap frequency, the superconducting materials become lossy due to the absorption caused by the breaking of Cooper pairs into quasiparticles.

\vspace{6mm}
\noindent\textbf{Current engineering}

\begin{figure*}
    \centering
    \includegraphics[width=\textwidth,trim=16 0 4 16,clip]{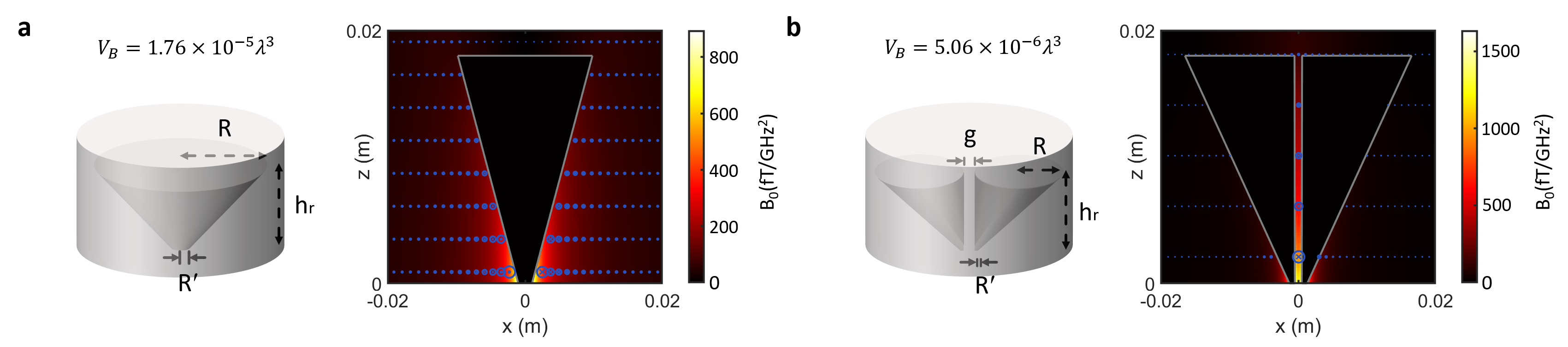}
    \caption{\textbf{Inverse-tapered reentrances for current crowding.} \textbf{a}, A tapered reentrant cavity ($f = 1.41$~GHz). The radius of the reentrance top and bottom are $R=1$~cm and $R'=1$~mm, respectively. Reentrance height is $h_r = 1.8$~cm. \textbf{b}, A tapered doubly reentrant cavity ($f = 2.12$~GHz). $R=8$~mm, $R'=0.8$~mm, $h_r = 1.8$~cm, and the gap between two reentrances is $g=2$~mm.}
\end{figure*}

\noindent Figure 3\textbf{a} shows the squeezed TM$_{010}$ mode on the $xz$plane. The mode has a uniformly strong magnetic field around the sidewall. The tangential magnetic field on the conductor is proportional to the surface current, which induces ohmic loss. On the other hand, $V_B$ strongly depends on the maximum magnetic field in Eq.~(\ref{modeVolume}), where the dipole is assumed to be. 

We consider the cylindrical cavity as an $LC$ oscillator where the surface charge (current) couples to the electric (magnetic) fields of the TM$_{010}$ mode; $\rho_s = \hat{\textbf{\text{n}}}\cdot\epsilon_0\vec{\textbf{\text{E}}}$ and $J_s = \hat{\textbf{\text{n}}}\times\vec{\textbf{\text{H}}}$, where $\rho_s$ and $J_s$ are the surface charge and current density, and $\hat{\textbf{\text{n}}}$ is the surface normal unit vector. For the TM$_{010}$ mode, charge oscillates between two parallel plates and current flows through the side wall $\pi/2$ out of phase as the electric and magnetic fields (Fig. 3\textbf{a}).

To strengthen the local magnetic field, we must locally increase the current density. The split-mode cavity in Fig.~3\textbf{b} achieves high current density using two reentrances on the top and bottom \cite{le2016towards}. Because the circumference of the reentrances is smaller than the circumference of the cylindrical cavity, the current density is higher on the reentrance surfaces. In addition, the increased capacitance between the top and bottom plates stores more charge, resulting in a larger current. Compared to the same height cavity shown in Fig.~2\textbf{a}, the mode volume decreased from $0.140 \lambda^3$ to $6.76\times 10^{-3} \lambda^3$. The increased $L$ and $C$ decrease the resonance frequency from $5.74$ GHz to $3.39$ GHz. The reentrant cavity \cite{carter2007calculation} or the loop-gap cavity \cite{park2016randomized} can be understood in the same way. Figure 3\textbf{c} shows a reentrant cavity with reentrance radius $R=5$~mm and height $h_r = 1.8$~cm. The cavity has $V_B=1.44\times 10^{-3}\lambda^3$ at $f=2.23$~GHz.

On the other hand, a doubly reentrant cavity~\cite{creedon2015strong} has two reentrances and a strong magnetic field between them. Figure 3\textbf{d} shows a doubly reentrant cavity with $R=3$~mm and the reentrance gap $g=4$~mm, further shrinking the mode volume to $V_B=4.03\times10^{-4}\lambda^3$ at $f=3.21$~GHz. One way to understand the mode of the cavity is to combine the modes of two single reentrant cavities, which form symmetric and antisymmetric modes \cite{goryachev2014high}. The antisymmetric mode has a strong magnetic field between the posts induced by two opposite direction currents. Alternatively, the mode can be viewed as a current-engineered TM$_{110}$ mode. Because it originated from the TM$_{110}$ mode, its resonance frequency ($3.21$~GHz) is higher than that of TM$_{010}$-mode derived reentrant cavity ($2.23$~GHz). (see Supplementary Note 1 for the discussion on the additional mode in a doubly reentrant cavity.) 

We show that current engineering also allows arbitrarily small $V_B$. Consider a reentrant cavity with a vanishingly narrow reentrance, that is, $R\rightarrow 0$. The capacitance is approximated to that of a cylindrical capacitor, $C\propto 1/\ln(a/R)$. The charge, proportional to the capacitance, moves through the reentrance. Applying the integral form of Ampere's law, the magnetic field on the reentrance surface diverges as $R\rightarrow 0$ with $(R\ln(a/R))^{-1}$ scaling (see the agreement with simulations in Supplementary Note 2). Similarly, the maximum magnetic field of the doubly reentrant cavity diverges as $\propto(R\ln(g/R))^{-1}$ for $g \gg R$ and $\propto 1/g$ for $R \gg g$ (see Methods). 

We can further reduce $V_B$ using the current crowding. In our framework, the capacitance of the reentrance accumulates charge. The charges oscillate through a narrow reentrance, producing a strong magnetic field. Thus, this design principle motivates a large capacitance and a narrow conducting channel. 

Figure 4 shows the modified reentrant cavity design, where we increased the current density by inverse-tapering the reentrance(s). This inverse tapering combines the benefits of a large capacitance with a reentrance top and current crowding. The modified reentrant cavity (Fig. 4\textbf{a}, $1.76\times10^{-5}\lambda^3$) has two orders of magnitude smaller $V_B$ than the non-tapered counterpart (Fig. 3\textbf{c}, $1.44\times10^{-3}\lambda^3$). Similarly, the doubly reentrant cavity also shows a two orders of magnitude reduction in mode volume from $4.03\times10^{-4}\lambda^3$ (Fig. 3\textbf{d}) to $5.06\times10^{-6}\lambda^3$ (Fig. 4\textbf{b}). 

Both designs also achieve arbitrarily small $V_B$. Consider a tapered reentrant cavity with fixed top radius $R$ and variable bottom radius $R'$. When we reduce $R'$ by more tapering, the capacitance does not decrease significantly as that of the non-tapered one because the top dominates the capacitance. Thus, the maximum magnetic field of the tapered reentrant cavity scales as $1/R'$. Similarly, the magnetic field of a doubly reentrant cavity $\propto 1/R'$ for $g \gg R'$ and $\propto 1/g$ for $R' \gg g$. In addition to the better scaling, the proportionality constant in both cases is larger than the non-tapered ones by increased capacitance.

\vspace{6mm}
\noindent\textbf{Magnetic field expulsion}

\begin{figure*}
    \centering
    \includegraphics[width=\textwidth,trim=16 4 0 8,clip]{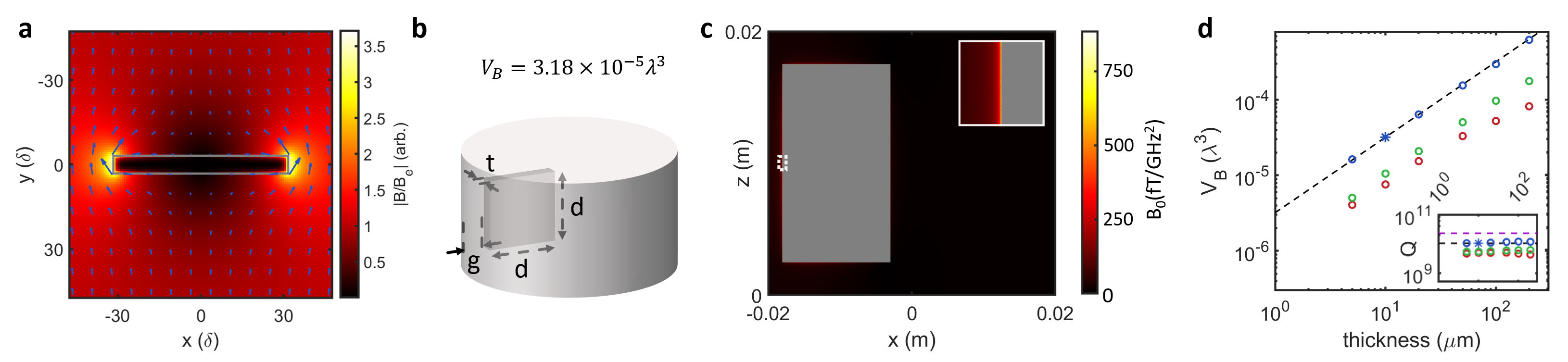}
    \caption{\textbf{Magnetic field expulsion.} \textbf{a}, Magnetic field expulsion by a thin metallic plate. The field distribution is calculated assuming the quasi-static limit, where the size of the plate is much smaller than the wavelength of the external field ($B_e$). The magnetic field near the edge of the plate is enhanced compared to the applied. \textbf{b}, A cylindrical cavity with a thin metallic plate of square shape, with the thickness $t$ and the length $d$. The side of the plate is electrically floated and has a gap $g$ with the cavity wall. \textbf{c}, The magnetic field distribution of TM$_{010}$-like mode. $d=1.5$ cm $g=2$ mm and $t=10~\mu$m for the metallic plate inserted. (Inset) close-up view of dotted rectangle region. \textbf{d}, $V_B$ vs. thickness of the plate. (blue) $d = 1.5$ cm, $g = 2$ mm, (green) $d = 1.9$ cm, $g = 2$ mm, (red) $d = 1.5$ cm, $g = 100~\mu$m. The blue asterisk represents the cavity in \textbf{c}. (dashed line) $V_B/\lambda^3 = (3.2~\mu m^{-1})\cdot t$. $\vec{\textbf{\text{r}}_\text{e}}$ is chosen for maximum $\vec{\textbf{\text{B}}}(\vec{\textbf{\text{r}}_\text{e}})$ restricted with half the thickness away from the plate. The inset shows $Q$ of the cavities, which changes negligibly on $t$. As a reference, $Q$ of the plain cylindrical cavity is marked with a purple dashed line.}
\end{figure*}

\noindent Lastly, we use magnetic field expulsion for $V_B$ reduction. The magnetic field expulsion follows from Faraday’s law, which shows that good conductors ($\sigma \gg \omega\epsilon_0$) and superconductors expel magnetic field: 
\begin{equation}
\begin{split}
\vec{\nabla}\times\vec{\textbf{\text{E}}}=-j\omega\vec{\textbf{\text{B}}} \xrightarrow{\vec{\textbf{\text{E}}}=0} \vec{\textbf{\text{B}}}=0.
\end{split}
\end{equation}
More specifically, the magnetic fields in the quasistatic limit follow;
\begin{equation}
\begin{split}
    \nabla^2\vec{\textbf{\text{B}}} = \delta^2\vec{\textbf{\text{B}}},
\end{split}
\end{equation}
where $\delta$ is the penetration depth. For superconductors, $\delta=\lambda_L$ is the London penetration depth (Meissner effect), and expulsion occurs at frequencies below the superconducting gap (e.g. Nb: $725$~ GHz). For normal conductors, $\delta = (\pi f\sigma\mu)^{-1/2}$ is the skin depth.

Figure 5\textbf{a} shows the magnetic field expulsion of a thin plate in the quasistatic limit. The external magnetic field circumvents the conductor resulting in a stronger magnetic field on the edge. The more elongated the conductor, the stronger the magnetic field on the edge. We can express the magnetic field on the surface of a conductor by the demagnetization factor ($N$) \cite{pozar2011microwave, tinkham2004introduction, jackson2007classical, osborn1945demagnetizing},
\begin{equation}\label{demagnetization}
\begin{split}
    B_{\text{surf}} = \frac{B_0}{1-N},
\end{split}
\end{equation}
where $B_0$ and $B_{\text{surf}}$ are the external magnetic field and the field on the surface, respectively (see Methods). For an infinitesimally thin conducting plate, $N\rightarrow 1$, and the surface field diverges. Physically, the penetration depth smears out the geometric structure, ending up with finite field strength. 

Figures 5\textbf{b} and 5\textbf{c} show a cavity design after inserting a thin square plate with $d = 1.5$~cm and thickness $t = 10~\mu$m, in the cylindrical cavity ($h=2$~cm). The gap between the cylindrical wall and the plate is $g=2$~mm. We assume that the plate is electrically disconnected from the cylinder and floats in the simulation to avoid complications from a dielectric. In a real device, one can prepare a device on a low-loss, low-permittivity dielectric substrate to physically support the plate. As a plate, one can attach a metallic foil to the substrate or deposit a film of metal on the substrate. (see Methods - Dielectric substrate effect). With this minimal modification, the mode volume of the cavity reaches $V_B = 3.18\times10^{-5}\lambda^3$. 

Figure 5\textbf{d} shows the dependence of the mode volume on the thickness of the inserted plate (blue). We find that $V_B$ is linearly proportional to the thickness of the plate. An important observation is that $Q$ does not decrease when $V_B$ is reduced by thinning the plate (Fig. 5\textbf{d}, inset). 

We can further reduce the mode volume by increasing $d$ or decreasing $g$. The green markers in Fig. 5\textbf{d} show the mode volume for $d=1.9$~cm with the same $t$ and $g$. Similarly, decreasing the gap to $g=100~\mu$m reduces $V_B$ (red). From the inset, we find that these modifications also reduce $Q$. However, in all cases, the decrease in $t$ does not degrade $Q$.

\vspace{6mm}
\noindent\textbf{Cavity comparison}

\noindent We first define a material-independent metric for the loss of a cavity. The geometric factor of a cavity is
\begin{equation}
\begin{split}
    G = \frac{\omega\mu_0\int|\vec{\textbf{\text{B}}}|^2dV}{\int|\vec{\textbf{\text{B}}}|^2dS}.
\end{split}
\end{equation}
We can express quality factors of a hollow metallic cavity with $G$ and surface resistance ($R_s$) of the metal;
\begin{equation}
\begin{split}
    Q = \frac{G}{R_s}.
\end{split}
\end{equation}

\begin{figure*}
    \centering
    \includegraphics[width=\textwidth,clip]{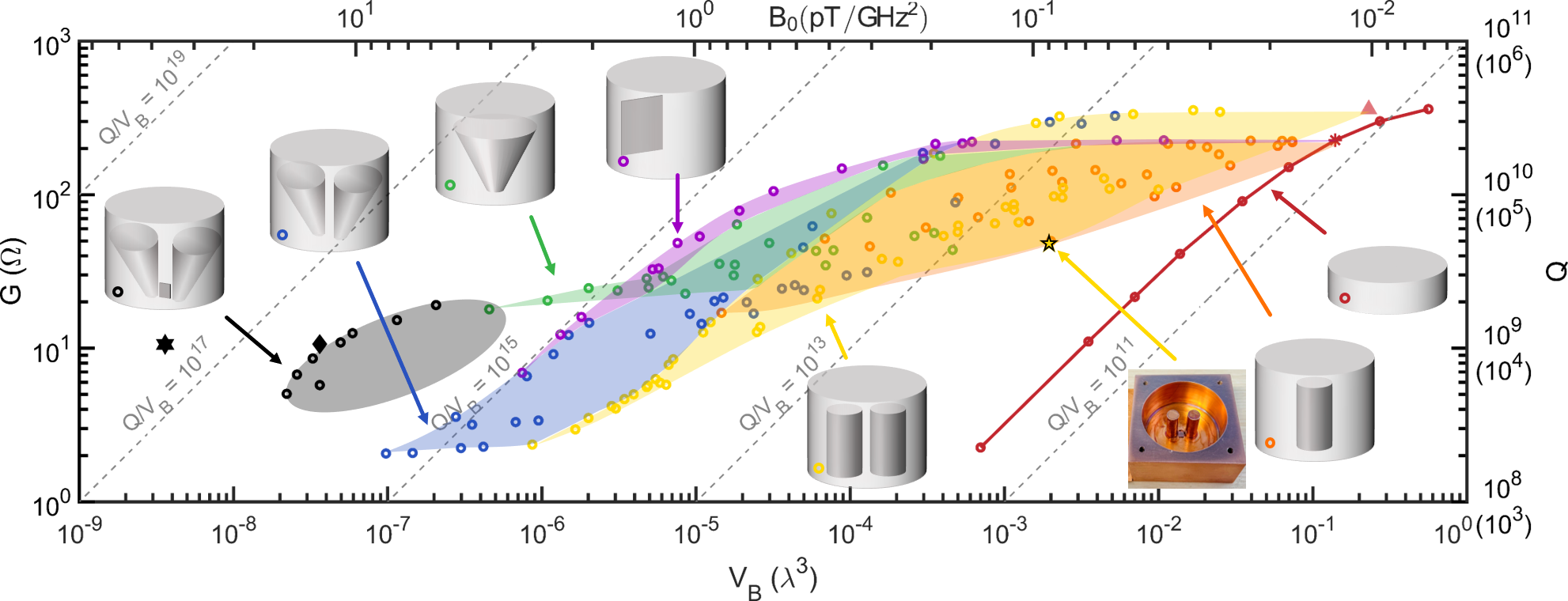}
    \caption{\textbf{Quality factors vs. mode volumes} of cavity designs. Left axis: geometric factors. Upper axis: single-photon magnetic field. $Q$ is calculated assuming Nb is used ($R_s = 10$~n$\Omega$). The values in the parenthesis are for Cu ($R_s = 1$~m$\Omega$). The mode volume (bottom axis) is normalized to the diffraction-limited volume ($\lambda^3$). Each color of the marker represents different cavity designs, as illustrated. Each marker is the simulated value of different cavity dimension parameters, and the shaded region is the trade-off space the design can access with the fabrication assumptions. The markers and shades are representative (see main text). The yellow star shows the experimentally demonstrated cavity. Our analysis shows that the combination of current engineering and field expulsion methods can achieve $V_B = 3.6 \times 10^{-9}\lambda^3$ (black star).}
\end{figure*}

Figure 6 summarizes the trade-off between $G$ and $V_B$ for cavity designs. The right $y$ axis shows the expected $Q$ for the Nb (Cu) cavity with $R_s = 10$ n$\Omega$ ($1$ m$\Omega$). Each color represents different cavities with illustrations. We fixed $h$ for all designs except the cylindrical cavity design. The minimum size of the reentrances is $100~\mu$m, and the minimum thickness of the plates is $10~\mu$m. The magnetic field for the calculation of $V_B$ is probed $50~\mu$m away from the reentrances or $5~\mu$m away from the edge of the plate considering the placement of a dipole. We emphasize that the markers and shades are representative, but do not limit the accessible $Q$ and $V_B$. Higher $Q$ or smaller $V_B$ are, in principle, possible with more aggressive assumptions on the cavity fabrication or operation frequencies. 

The red circles show the simulated value of cylindrical cavities. As $h$ decreases, $V_B$ decreases proportionally. A longitudinally squeezed cavity with $h=100~\mu$m reaches $V_B=7.00\times10^{-4}\lambda^3$. $Q$ is nearly unchanged with changing $h$ for $h\gg r$ when the side wall dominates the loss, while $Q$ decreases proportionally with $h$ for $h\ll r$ due to the dominant loss from the top and bottom. Thus, longitudinal squeezing decreases both $V_B$ and $Q$, with $Q/V_B\sim 3\times10^{11}\lambda^{-3}$ (Nb) unchanged for $h\ll r$. 
Current engineered cavities achieve not only smaller $V_B$ but also higher $Q/V_B$. Starting from the cylindrical cavity with $h=2$~ cm (red asterisk), we introduce a reentrance (orange). For a thin and short reentrance, $Q$ barely changes with reduced $V_B$. However, to further reduce the mode volume, given the minimum radius $R=100~\mu$m, we need taller reentrances to increase the capacitance. When the sidewalls dominate the loss (small $h_r$), increasing $h_r$ does not decrease $Q$. When the reentrance dominates the loss (large $h_r$), increasing $h_r$ decreases $V_B$ with marginal improvement in $Q/V_B$. 

Doubly reentrant cavities (yellow circles) also have a $V_B$-$Q$ trade-off, starting from the TM$_{110}$ mode (pink triangle). The smallest $V_B\approx8.68\times10^{-7}\lambda^3$ is an order of magnitude smaller than that of the reentrant cavities ($V_B\approx1.46\times10^{-5}\lambda^3$). The inverse-tapered reentrant (green circles) and doubly reentrant cavities (blue circles) have similar behavior but with an additional order of magnitude smaller $V_B=4.56\times10^{-7}\lambda^3$ and $V_B=9.39\times10^{-8}\lambda^3$, respectively. The inverse-tapered cavities include the trade-off space of non-tapered cavities, which we did not include for visibility.

Field expulsion cavities (purple circles) can reach $V_B=3.57\times 10^{-4}\lambda^3$ with a negligible reduction in $Q$. The flat $V-Q$ trade-off is obtained by thinning the inserted plate. A further reduction of $V_B$ is possible with increasing $d$ or decreasing $g$, and the minimum $V_B = 7.45\times 10^{-7}\lambda^3$.

Lastly, we present a cavity design that combines current engineering and field expulsion. In the design, we inserted a thin plate into the double reentrant, inversely tapered cavity (black circles). The highest $Q/V_B$ approaches $3\times10^{16}\lambda^{-3}$ (black diamond), which is five orders of magnitude larger than the bare cylindrical cavity. If the inserted plate is thinned to $1~\mu$m, the mode volume will reach $V_B = 3.6\times 10^{-9}\lambda^3$ (black star). 

We also emphasize that the mode volume in units of $\lambda^3$ can be much smaller at lower operation frequencies. In this analysis, we consider $f = 1\sim10$ GHz for the applications below, but by operating at $f \approx 10$ MHz, one could easily achieve $V_B<10^{-11}\lambda^3$ and $Q/V_B>10^{20}\lambda^{-3}$ with a minimum feature size of $100~\mu$m and a plate thickness of $1~\mu$m. Note that $Q/V_B=10^{20}\lambda^{-3}$ corresponds to Purcell enhancing an emitter decaying slower than the age of the universe, to decay within a few tens of milliseconds.

\vspace{6mm}
\noindent\textbf{Experimental Validation}

\noindent We ensured the mode engineering method and the validity of the analysis by fabricating a reentrant cavity and characterizing the mode volume with nitrogen-vacancy (NV) centers in diamond. NV centers are atomic defects in diamond lattice with a long room-temperature coherence time of up to 1.5 milliseconds~\cite{herbschleb2019ultra}. Moreover, one can measure the spin states of NVs because they have different photoluminescence intensities due to the intersystem crossing to the singlet. We used an ensemble of NV centers as an MW magnetic field sensor using the optically detected magnetic resonance method.

We designed the cavity based on Fig. 3\textbf{d} with the modification of height ($h=1.9$ cm) to be resonant with NV centers ($f=2.87$ GHz). We placed an electronic-grade $3\times3\times 0.5$ mm$^3$ diamond with an NV density of 300~ppb, between two reentrances where the magnetic field is strong (see Methods - Cavity fabrication). We strongly drove (25 dBm) spins and measured the fluorescence. Fluorescence oscillates with driving time (Rabi oscillation) and the signal is fit to the sum of three damped oscillations corresponding to the three groups of NVs that orient differently (see Supplementary Note 5 and Fig.~S6). The oscillation frequency corresponding to the resonant NV ensemble was $2.929$ MHz, corresponding to $V_B = 1.75\times 10^{-3} \lambda^3$. This value is close to the numerically simulated value of $1.95\times 10^{-3} \lambda^3$. We also characterized the cavity loss with a vector network analyzer. The measured $f = 2.871$ GHz and $Q = 2,421$ are similar to the simulated value of $2.864$ GHz and $3,169$, respectively (the simulated $Q$ is based on the surface resistance room temperature Cu, $R_s = 15.1$ m$\Omega$).

With this preliminary demonstration, we confirm that the simulation and experiment agree on three key parameters: 1) the mode volume ($V_B$), 2) the quality factor ($Q$), as well as 3) the resonance frequency ($f$). This measurement result is shown in Fig. 6 as a yellow star.

\section{Discussion}

\noindent Extreme values of $Q/V_B$ with our cavity designs open up new possibilities in the control of magnetic light-matter interactions, including (i) microwave photon-spin coupling, (ii) magnetic circuit QED, (iii) ultra-precision metrology \cite{ye2008quantum}, (iv) dark-matter searches \cite{kahn2016broadband,garcon2019constraints}, etc. We briefly elaborate on the first two examples (see also Methods for the calculations). 

(i) Solid-state spin qubits, such as nitrogen vacancy centers, have shown a coherence time ($T_2^*$) of 1.5 ms \cite{herbschleb2019ultra}. Coupling a single NV center to the cavity indicated in Fig. 6 as a black diamond, at $10$ GHz, would show a high cooperativity $C=4g^2/\kappa\gamma^*=12.5$, where $g$ is the coupling rate, $\kappa$ is the cavity dissipation rate and $\gamma^*$ is the dephasing rate. The interaction can be greatly increased with spin ensembles or magnons by the number of spins; $C=4Ng^2/\kappa\Gamma$ where $N$ is the number of spins, and $\Gamma$ is the ensemble dephasing rate.  Spin ensemble-based technologies, such as optical transduction \cite{fernandez2015coherent,everts2019microwave,bartholomew2020chip}, can achieve a high-cooperativity system with our designs. However, the single-spin operation enables full quantum nonlinearity with a wide range of applications, including high-fidelity single MW-photon optical transduction~\cite{kurokawa2022remote} for quantum networks and thresholdless~\cite{kobayashi1982novel}, single-spin~\cite{mckeever2003experimental} ultralow noise masers~\cite{breeze2018continuous}. The number of spins in our calculation is twelve orders of magnitude smaller than the early efforts \cite{kubo2010strong, schuster2010high, creedon2015strong} and six orders of magnitude smaller than the recent state of the art with the lumped-element / 3D hybrid approach with $N\sim1.5\times10^6$ \cite{bienfait2016magnetic}.  

(ii) The flux qubit family (RF-SQUID, 3-junction flux qubit, and fluxonium) can be coupled to the cavity. $6~\mu$m$\times5~\mu$m loop used in \cite{yan2016flux} can magnetically couple with the cavity with $g\sim 2.33$~MHz at $5$ GHz, achieving a strong coupling regime. In addition, the qubit will be located at the magnetic field antinode, where the electric field is minimal, potentially introducing negligible cavity loss from two-level systems in substrate dielectrics. 

%% diameter: 5 um line width: 1 um, thickness : 250 nm
%% L ~ 5 pH
%% Ip ~ 100 nA
%% A ~ 7.85E-11 m^2
%% LJ << L
%% m ~ 2E-15*A/L ~ 3E-14    (phi = I*L, m = I*A, phi = 2E-15)
%% B_0 = 325 pT = 3.25E-10     (5 GHz)
%% B0*m/h ~ 14.71 GHz

% Ip = 1E-7;
% A  = 30E-12;
% B0 = 515E-12 T
% B0*m = 2.33 MHz

\section{Conclusion}

\noindent In conclusion, we described design methods to reduce the magnetic mode volume of electromagnetic cavities. The resulting mode volume of all three methods can be arbitrarily small, only limited by the material's penetration depth or, practically, the fabrication resolution. We benchmarked different cavity designs in terms of mode volume and loss. The geometric factor, a material-independent metric for loss, only decreases ten times for more than a million times smaller mode volume (black diamond in Fig. 6). This corresponds to $Q=1.06\times10^9$ with $V_B=3.62\times10^{-8}\lambda^3$ for superconducting niobium. This opens the possibilities of quantum applications such as cavity quantum magneto-dynamics (cQMD) with spins or superconducting artificial atoms, and magnetic nonlinear devices.

Future work should modify the designs for ease of fabrication, operational frequency, multimode operation, or application-dependent figures of merit (see some examples in Supplementary Notes 3 and 4). For example, the figures of merit are $1/V_B$ (Purcell effect in BE), $1/\sqrt{V_B}$ (vacuum Rabi splitting in BE), or $Q/\sqrt{V_B}$ (vacuum Rabi splitting in BC) \cite{choi2017self}, where BE and BC stand for bad emitter and bad cavity regime. Nonlinear (magnetic) optical applications benefit from a high magnetic field over the volume of nonlinear medium with the figure of merit $Q^2/V_\text{Kerr}$ (nonlinear bistability) or $QV_M/V_B^2$ (Kerr blockade) \cite{choi2017self}. Moreover, cascaded cavities coupled with quantum emitters have richer dynamics with advantages over the single cavity system~\cite{choi2019cascaded}. Furthermore, certain applications will benefit from other materials, such as high-T$_\text{c}$ superconductors for liquid nitrogen temperature operations.

%\bibliographystyle{naturemag}
%\bibliography{references}

\clearpage
\noindent\textbf{Methods}
\footnotesize

\noindent\textbf{Numerical simulation of cavities.} All numerical results are obtained using the finite element method (FEM) with commercial software (COMSOL). We used tetrahedral meshes with a maximum size of $1\sim2~$mm. The minimum mesh size is chosen to be less than half the minimum feature size (e.g., thickness of the plate in Fig. 5\textbf{c}). The maximum element growth rate is set to $1.35\sim 1.45$ with a curvature factor of $\sim 0.4$ and a narrow region resolution of $\sim 0.7$.

\vspace{5mm}
\noindent\textbf{TM$_{010}$ mode volume.} 
\begin{equation}
\begin{split}
    \int dV B^2/\mu_0 &= \int dV \epsilon_0 E^2 \\
    &= 2\pi h \epsilon_0 |E_0|^2 \int_0^a r|J_0(j_{01}r/a)|^2 dr \\
    &= 2\pi h \cdot a^2 \cdot \epsilon_0 |E_0|^2 \int_0^1 x|J_0(j_{01}x)|^2 dx \\
    &\approx 0.135\cdot 2\pi h \cdot a^2 \epsilon_0|E_0|^2.
\end{split}
\end{equation}
\begin{equation}
\begin{split}
    V_B &= (0.135 \cdot 2\pi h \cdot a^2)\cdot(\frac{a}{j_{01}})^2\cdot \frac{\omega^2\mu_0\epsilon_0}{|J_0'(j_\text{max})|^2} \\
    &\approx \frac{0.135}{0.339}\cdot 2\pi h \cdot (\frac{j_{01}}{k})^2\\
    &\approx 0.0635 \cdot |j_{01}|^2 \cdot h \cdot \lambda^2\\
    &\approx 0.366 \cdot h \cdot \lambda^2
\end{split}
\end{equation}
where $j_\text{max}\in[0,j_{01})$ maximizes $|J_0'(j_\text{max})|^2$.

\vspace{5mm}
\noindent\textbf{$V_B$ scaling of doubly reentrant cavities.} The capacitance of two-wire transmission line (per unit length) is $C=\pi\epsilon_0/\cosh^{-1}\left((g+2R)/2R\right)\approx\pi\epsilon_0/\ln(g/R)$ for $g\gg R$. The currents proportional to the capacitance are uniformly distributed over the circumference of the reentrance in this limit. Thus, the maximum magnetic field scales as $(R\ln(g/R))^{-1}$. For $g\ll R$, $C\approx\pi\epsilon_0\cdot\sqrt{R/g}$. Current proportional to the capacitance creates a flux proportional to inductance $L\approx\pi\mu_0\cdot\sqrt{g/R}$. The scaling of $L$ and $C$ cancels each other, and the magnetic field scales as $\sim 1/g$ assuming that it is uniform across the gap. 

\vspace{5mm}
\noindent\textbf{The demagnetization factors.} The relation between $B_\text{surf}$ and $N$ shown in Eq.~(\ref{demagnetization}) needs calculation of $N$. An ellipsoid in a uniform magnetic field has a uniform demagnetization field~\cite{jackson2007classical}, and $N$ can be unequivocally defined; $N=1-E(\sqrt{1-b^2/a^2})\cdot c/b$, where $a\geq b\gg c$ are half the length of the principal axes, and $E$ is the complete elliptic integral of the second kind. (See \cite{osborn1945demagnetizing} for the full analytic expression in general cases.) We find as $a=b$ and $c/b\rightarrow 0$, that $N \approx 1-c/b$ and $B_\text{surf}\approx b/c \rightarrow \infty$. We can extend Eq.~(\ref{demagnetization}) to non-ellipsoidal cases by defining an effective demagnetization factor $N_\text{eff} = 1-B_0/B_\text{surf}$. It can be easily shown that an infinitesimally thin superconducting plate with $\lambda_L\rightarrow 0$ has $N_\text{eff}\rightarrow 1$ from the magnetic Poisson's equation \cite{jackson2007classical}.

\vspace{5mm}
\noindent\textbf{Dielectric substrate effect.} In the simulations of the field-expulsion cavities, we assumed that the inserted plate is floating in the cavity. For the fabrication of the device, one can attach a foil or deposit a thin film on a low-loss, low-permittivity dielectric substrate. At the single-photon level, Heat exchanger method (HEM) sapphire is an option for the substrate and has shown a loss tangent of $1.5\times 10^{-8}$ (bulk) and $0.9\times 10^{-3}$ (surface, 3nm interface)~\cite{read2022precision}. In the linear regime with a large photon number, the loss is two orders of magnitude smaller~\cite{luiten1993ultrahigh} due to the saturation of two-level systems.

We simulated the field-expulsion cavity in Fig. 5\textbf{b}, with a 100~$\mu$m-thick, 2 mm-wide sapphire as a substrate, which extends to the bottom of the cavity to support the foil (see Supplementary Note 6). The simulation showed a negligible change in resonance frequency ($\sim0.34$\%) and mode volume ($\sim7.5$\%), and additional dielectric loss at the single-photon level, was characterized as $Q_\text{dielectric}\sim 1.45\times 10^{10}$. Thus, the additional loss from the sapphire substrate is smaller than the loss from the metallic parts even for the superconducting case ($Q_\text{s.c.}\sim 1.05\times 10^{10}$). For simplicity, we ignored the dielectric substrates in the cavities.

\vspace{5mm}
\noindent\textbf{Cavity fabrication.} We began by milling a high-purity C101 copper block and polishing it with sandpaper to ensure a smooth surface. We then cleaned the inner surfaces of the cavity using soldering flux. An SMA connector is soldered on the top plate, with its center pin exposed to the inside. By adjusting the length of this pin, we modified the external coupling in our experiment. A small hole at the bottom provides optical access for initializing and probing NV centers in diamonds. It is assured that the bottom wall of the cavity was thick enough to prevent microwave leakage losses through the hole. For more information, please refer to Supplementary Note 5 and Fig.~S5.

\vspace{5mm}
\noindent\textbf{Coupling rate calculation.} For solid-state spin qubits, we used the gyromagnetic ratio of electronic spin $\gamma = 28$~GHz/T. At $10$~GHz, $V_B = 3.62\times10^{-8}\lambda^3$ corresponds to $B_s=2.06$~nT and $g = 57.7$~Hz.

For a superconducting flux qubit, we calculate the magnetic transition dipole moment, $m_0$. Near the sweet spot, the computational basis of a flux qubit is a symmetric and antisymmetric combination of clockwise and anticlockwise circulating current (equivalent to $n$ and $n+1$ flux quantum in the superconducting loop): $\ket{0} = \sqrt{1/2}(\ket{\CircleArrowright} + \ket{\CircleArrowleft})$ and $\ket{1} = \sqrt{1/2}(\ket{\CircleArrowright} - \ket{\CircleArrowleft})$. The transition dipole moment is
\begin{equation}
\begin{split}
    m_0 = \bra{0}\hat{m}\ket{1} &= \frac{1}{2}(\bra{\CircleArrowright}\hat{m}\ket{\CircleArrowright} - \bra{\CircleArrowleft}\hat{m}\ket{\CircleArrowleft}) \nonumber\\
    &= A\cdot\bra{\CircleArrowright}\hat{I}\ket{\CircleArrowright},
\end{split}
\end{equation}
where $\hat{I}$ is the current operator and $A$ is the area of a superconducting loop. Assuming a $6~\mu$m$\times5~\mu$m loop and circulating current $\bra{\CircleArrowright}\hat{I}\ket{\CircleArrowright}=100$~nA, used in \cite{yan2016flux}, $m_0=3\mu$A$\cdot\mu$m$^2$. Thus, $g = B_s\cdot m_0/h = 2.33$~MHz at $f=5$~GHz ($B_s$ of the cavity marked as the black diamond in Fig. 6.)
\normalsize

\vspace{6mm}
\noindent\textbf{Data availability}

\noindent The data supporting the plots in this paper and other findings of this study are available from the corresponding authors upon reasonable request.

\vspace{6mm}
\noindent\textbf{Acknowledgement}

\noindent We thank Kevin O'Brien, Karl Berggren, Qing Hu, William D. Oliver, Owen Miller, Richard Averitt, Eugene Demler, Youngkyu Sung, Isaac Harris, Matt Trusheim, Don Fahey, Jonathan Hoffman, Daniel Freeman, Shantanu Rajesh Jha, Max Hays, Shoumik Chowdhury, and Kyle Serniak for fruitful discussion. This work was partially supported by the Defense Advanced Research Projects Agency (DARPA) DRINQS (HR001118S0024) program and MITRE Quantum Moonshot. H.C acknowledges the Claude E. Shannon Fellowship and the Samsung Scholarship. 

This version of the article has been accepted for publication, after peer review. This is not the Version of Record and does not reflect post-acceptance improvements, or any corrections. The Version of Record is available online at: \href{https://doi.org/10.1038/s42005-023-01224-x}{https://doi.org/10.1038/s42005-023-01224-x}.

\vspace{6mm}
\noindent\textbf{Author Contributions}

\noindent H.C. conceived and simulated the design methods. H.C. and D.E. analyzed and discussed the results. H.C. wrote the manuscript with the assistance of D.E.

\vspace{6mm}
\noindent\textbf{Competing interests}

\noindent Authors declare no conflicts of interest.

\vspace{6mm}
\noindent\textbf{Additional information}

\noindent Supplemental information is available for this paper.

\normalsize

\clearpage

\end{document}